\documentclass[preprint2]{aastex}

\begin{document}

\title{FAR ULTRAVIOLET IMAGERY OF THE EDGE-ON SPIRAL GALAXY NGC~4631}

\author{Andrew M. Smith\altaffilmark{1}, Nicholas R. Collins\altaffilmark{2}, 
William H. Waller\altaffilmark{3}, 
Morton S. Roberts\altaffilmark{4}, \\ Denise A. Smith\altaffilmark{5}, 
Ralph C. Bohlin\altaffilmark{5}, K.-P. Cheng\altaffilmark{6}, 
Michael N. Fanelli\altaffilmark{7}, 
Susan G. Neff\altaffilmark{1}, Robert W. O'Connell\altaffilmark{8},
Ronald A. Parise\altaffilmark{9}, 
Eric P. Smith\altaffilmark{1}, 
and Theodore P. Stecher\altaffilmark{1}}

\altaffiltext{1}{Laboratory for Astronomy and Solar Physics, Code 681, NASA/GSFC, Greenbelt MD 20771}
\altaffiltext{2}{Raytheon STX Corporation, Code 682, NASA/GSFC, Greenbelt MD  20771}
\altaffiltext{3}{Raytheon ITSS Corporation, Harvard-Smithsonian Center for Astrophysics and Tufts University. Current Address: Department of Physics and Astronomy, Tufts University, Medford MA 02155} 
\altaffiltext{4}{National Radio Astronomy Observatory, Edgemont Road, Charlottesville, VA 22903}
\altaffiltext{5}{Space Telescope Science Institute, 3700 San Martin Drive, Baltimore, MD 21218}
\altaffiltext{6}{Department of Physics, California State University, Fullerton, 800 N. State College Ave., Fullerton, CA 92634}
\altaffiltext{7}{Department of Physics, University of North Texas, P.O. Box 305370, Denton, TX 76203}
\altaffiltext{8}{Department of Astronomy, University of Virginia, P.O. Box 3018, Charlottesville, VA 22903}
\altaffiltext{9}{Computer Sciences Corp., Code 686.9, NASA/GSFC, Greenbelt MD  20771}

\begin{abstract}

Far ultraviolet ($FUV$) imagery of the edge-on, Sc/SBd galaxy,
NGC~4631 reveals very strong $FUV$ emission, resulting from active
star formation, uniformly distributed along the galactic mid-
plane. Multi-band imagery, \ion{H}{1} and
\ion{H}{2} position-velocity curves and extinction considerations all imply
that the emission is from the outer edges of the visible galaxy. The
overall $FUV$ morphology of this edge-on disk system is remarkably
similar to those of the so-called ``chain galaxies'' evident at high
redshift, thus suggesting a similar interpretation for at least some
of those distant objects.  $FUV, U, B$ and $V$ magnitudes, measured
for 48 star forming regions, along with corresponding $H\alpha$ and
$H\beta$ measurements are used to construct diagnostic color-color
diagrams.  Although there are significant exceptions, most of the star
forming regions are less massive and older than 30~Doradus. Comparison
with the expectations from two star formation models yields ages of
2.7 to 10~Myr for the instantaneous burst (IB) model and star
formation cut-off ages of 0 to 9~Myr for the continuous star formation
(CSF) model.  Interpreted in terms of the IB model the photometry
implies a total created mass in the 48 star forming regions of 
2.5$\times$10$^{7}$~M$_{\sun}$.  When viewed as resulting from constant star formation the
photometry implies a star formation rate of 0.33~M$_{\sun}$~yr$^{-1}$. These
results are compared to those derived from FIR and radio
observations. Corrections for $FUV$ emission reprocessed by interstellar
grains are estimated.

A large ring, $\sim$3~kpc in diameter, of 14 star forming regions is
concentrically located with an expanding \ion{H}{1} shell \citep{ran93} 
toward the eastern end of the galaxy. Our
observations imply that the shell may have been generated primarily by
supernovae arising from 5.3$\times$10$^{4}$ OB stars in a massive star forming
region beginning about 20~Myr ago, and that the presently observed $FUV$ bright
emission is due to second generation stars. 
 
\end{abstract} 

 
\section{Introduction} 

The group of galaxies including NGC~4631 provides an outstanding
example of a galaxy interaction accompanied by intensive star
formation. Far Ultraviolet ($FUV$) imagery of the Sc/SBd galaxy
NGC~4631, recorded by the Ultraviolet Imaging Telescope (UIT),
exhibits very bright $FUV$ emission from sources corresponding to the
\ion{H}{2} regions cataloged by \citet{cri69} 
 along with substantial diffuse $FUV$ light. The observed
emission is, in our experience, extraordinary in that NGC~4631 is
observed nearly edge-on, and strong attenuation of $FUV$ light by dust
in the galactic disk would be anticipated. Such is the case for
NGC~891, another highly inclined galaxy and M82, an edge-on starburst
galaxy -- both of which show very little detectable $FUV$ emission.
(Images of NGC~891 and M82 are accessible at
http://archive.stsci.edu/astro/uit/index.html).  Together with the
$H\alpha$ imagery obtained by \citet{ran92}
the $FUV$ results signify widespread intensive star formation along
the observable disk of NGC~4631. The overall $FUV$ morphology is
remarkably similar to those of the so-called ``chain galaxies'' that
have been observed at high redshift \citep{cow95}.  
It may be that a significant fraction of these distant
$FUV$ sources is composed of highly inclined disk galaxies rather than
dynamically unstable chains of sub-galactic clumps.  See \citet{smi97} 
for further details.

Observations of the radio halo imply that the galactic-wide star
formation in NGC~4631 is sufficiently intense as to supply cosmic rays
to the halo through bursting supershells formed by supernovae and
stellar winds in the disk \citep{dah95,gol94a}.
Optical images reveal a
patchy distribution of dust rather than a well defined dust
lane. Thus, the extensive $FUV$ emission may result from the combined
dispersal of dust by these bursting supershells and by forces
generated through tidal interactions NGC~4631 and NGC~4656 
\citep{com78}.

There is also evidence for an inner disk or torus (r$\sim$50$\arcsec$) revealed
most clearly in CO(1-0) and CO(2-1) line emissions 
\citep{gol94b,sof89,sof90}. The radio continuum emission coincides spatially
with the CO emission and has a substantial non-thermal component 
\citep{dur82}.  
There also appears to be a region of enhanced star
formation near the center (R$<$360~pc) of NGC~4631 \citep{dur82,sof89} coinciding
closely with the  center of the galaxy as measured in the infrared by 
\citet{aar78}. More recently,  \citet{wan95} reported
the ROSAT detection of soft X-rays extending out to 8~kpc above the galactic
midplane. When interpreted in terms of the burst supershell model, this
observation implies substantial star formation near the center of NGC~4631.
Indeed, considering mostly the radio data, \citet{gol94b} 
view NGC~4631 as a galaxy with  a {\it mild} central
starburst.

Of interest is a supershell with a diameter of 3~kpc detected at 21~cm
by \citet{ran93}. The shell
corresponds to a distortion in the disk near the eastern end of the
galaxy appearing in all optical images and in a particularly obvious
way in both the $FUV$ and $H\alpha$ images. A preliminary explanation
for the shell is that it is produced by the combined effects of
stellar winds and supernovae from OB associations containing
(1-3.5)$\times$10$^{4}$ OB stars \citep{ran93}.   
Alternatively, these authors suggest that the 
eastern cavity may have been punched into the galactic disk by a high
velocity cloud (HVC) which itself originated from the tidal stripping
of the galactic disk during a prior interaction with NGC~4656.  A
collision with a dwarf galaxy could have also produced the same
result.

In this paper we present the $FUV$ imagery of NGC~4631. Using these data together
with recently obtained optical imagery, we infer the star formation history of
the bright $FUV$ emitting associations including those circumscribed by the Rand
and van der Hulst super shell. The implications of these results on the
ultraviolet emission morphology, local and global star formation, and the
viability of the starburst shell and HVC impact hypothesis are discussed. In
all subsequent discussions, we place the galactic center at 
RA(2000)~=~12$^{h}$39$^{m}$ 39$\fs$8,  DEC(2000)~=~+32\arcdeg 48\arcmin
48\arcsec, adopt a galactic distance of 7.5~Mpc \citep{gol94a}  
and abbreviate the \ion{H}{2} regions observed by
\citet{cri69}  as CM(\#).
 
\section{Observations} 
 
NGC~4631 was observed by the UIT through the B1 filter
($\lambda_{eff}$~=~152.0~nm; $\Delta\lambda$~=~35.4~nm) for
1140~seconds on March 10, 1995 during the Spacelab ASTRO-2
mission. The instrument is completely described in 
\citet{ste92,ste97}.  $FUV$ images,
intensified and down-converted to visible wavelengths by a two stage
magnetically focussed intensifier, were recorded on IIa-O film,
digitized with a PDS microdensitometer, and reduced as described in
\citet{ste92,ste97}.
The plate scale of 1\arcmin~mm$^{-1}$ together with a
microdensitometer aperture of 0.021~mm resulted in a pixel size of
1.26\arcsec.  The FWHM of the point spread function was 3.4\arcsec~
over a circular 40\arcmin~ field-of-view.

Ground based images of NGC~4631 were also recorded in the standard  $U, B, V,$
and $R$ bandpasses and in $H\alpha$ and $H\beta$ line emission. The
observations are summarized in Table~1. The $FUV,$ $H\alpha$, $U, B,$ and
$V$-band images are shown in Figure~1.
 
%\begin{table}
%\dummytable\label{tab:1}
%\end{table} 

The absolute sensitivity of the UIT was determined by observations of stars the
spectra of which were measured by the IUE 
\citep{ste97}.  This procedure resulted in an uncertainty in the
absolute flux measurements of $\sim$10\%. Relative sensitivity throughout the
B1 passband was measured in the laboratory.  Relative transmissions anywhere in
the passband were accurate to 2\% at transmission levels above 1\% of the
maximum transmission. At 320~nm the instrument response is reduced from the
maximum response by more than 5~dex and at 330~nm by more than 8~dex.

The UIT image was corrected for geometric image (``S'') distortion according to
the formulation developed by \citet{gre94}.  
Each of the ground based images was then transformed to the same plate 
scale and orientation as the UIT image.

The absolute sensitivity of the $U, B, V,$ and $R$ band images was determined
using observations of the spectrophotometric standard star BD+33~2642. After
establishing the calibration using the spectrophotometry of 
\citet{sto77},  the $U, B,$ and $V$ magnitudes were cross checked
against the broad band measurements recorded by 
\citet{kle62}.  The values match within 5\%.

Astrometry of the broad band $U, B, V, R$ and $H\beta$ images was determined
using stars in common with the {\sl Digitized Guide Star Catalog} of \citet{las90}.  
The astrometric solution of the UIT image
was based on positions of star forming knots in common with the KPNO U-band
image yielding a positional uncertainty of $\pm$~0.74~arcseconds.  

Continuum subtraction in the $H\alpha$ image was performed using foreground
stars in both on-line and off-line images.  Observations of the planetary
nebula NGC~7027 were used for absolute calibration. Astrometry for the 
$H\alpha$ image was adopted from \citet{ran92}, 
and differs from the broad-band astrometry by no more than 1.1~arcseconds.

The on-line and off-line $H\beta$ images were separately calibrated using the
spectrophotometric standard star, Kopff~27 \citep{sto77}. After
absolute calibration, a scale factor of 1.22 was applied to the off-line image
to allow removal of continuum sources, i.e. star images from the field. After
continuum subtraction was performed, the flux of the UV-bright region in the
disk of NGC~4631, CM-67, as measured through a 21\arcsec~ diameter synthetic
aperture was 2.34$\times$10$^{-13}$ ergs s$^{-1}$ cm$^{-2}$.  This value is 6\%
greater than the equivalent measurement by \citet{roy91}.

\section{Morphology} 
 
Inspection of the V and B images shows dust silhouetted against the
red stars of the nuclear bulge. Most of the dust appears to be
concentrated near the galactic center; its morphology corresponds
roughly to the distribution of CO in an interior disk as measured by
\citet{gol94b},  \citet{sof90}, and of the dust as observed by 
\citet{bra95}, in 1.3~mm continuum emission.  Its
distribution is chaotic extending above the galactic midplane by as
much as 1.3~kpc. The B, V and $H\alpha$ images show that the cores of
some of the CM \ion{H}{2} regions are silhouetted against the dust
patches, implying that much of the bright $FUV$ emission originates
edgeward from the central 4~kpc of NGC~4631.

This impression is reinforced by the measured \ion{H}{1} and \ion{H}{2} 
velocity-position curves. The \ion{H}{1} velocity-position curves  
\citep{wel78,sof90} imply a
constant  \ion{H}{1} velocity at distances greater than $\sim$1\arcmin~  from
the  galactic center.  However, at these distances the $H\alpha$ data of 
\citet{cri69} show a linear increase in the
observed  velocity as a function of increasing distance from the galactic 
center. Such behavior would result if the \ion{H}{2} gas velocity were 
constant beyond $\sim$1\arcmin~  from the galactic center, and the  emission
was from gas located at an approximately constant galactocentric radius. Since the bright
FUV/\ion{H}{2} regions extend 24~kpc along the {\it entire} major axis of the imaged
galaxy, we deduce that most of  the \ion{H}{2} regions are located near the
galactic periphery. 

We also note that if the presently observed $FUV$ emission  regions were to be
further obscured by dust in an interstellar  gas of density 0.5~cm$^{-3}$, as
indicated by \citet{ran94} 21~cm  measurements of the NGC~4631
midplane, an effective  gas depth of  1.2~kpc would cause more than half of the
observed UV-bright  regions to be undetectable by the UIT. To obtain this
result, the  relationship 

\begin{displaymath}
\langle N(HI + H_{2}) / E(B-V) \rangle = 5.8\times10^{21}~atoms~cm^{-2}~mag^{-1} 
\end{displaymath}
\noindent
\citep{boh78}, determined from measurements 
along 75 lines-of-sight in the Galaxy, is assumed to be valid. We  do not
include the molecular gas density in our estimate since  this quantity is
unknown; its inclusion would further reduce the  effective gas depth. We also
assume that a detection would be at  the 2$\sigma$ level. Even though the dust
is observed to be patchy, the  fact that the entire major axis is delineated by
bright $FUV$  emission requires that many of the $FUV$ emission regions are 
near the visual periphery of the galaxy.  

Finally we note that peripheral \ion{H}{2} regions and $FUV$ emitting  sources
are commonplace in many Sc-Sd type galaxies, including  M101 
\citep{wal97a}, M74 \citep{cor94} and M83 \citep{boh83}. 

Generally, the $FUV$ emission coincides closely with the $H\alpha$
emission, much of which shows significant departures from the midplane
as defined by the symmetric distribution of \ion{H}{2} regions about
the major axis (see \citet{cri69}). If this
coincidence is related to the distribution of young, massive stars,
and contributions to the observed $FUV$ light from distant regions of
the galaxy are masked by the UV opacity of the displaced gas then star
formation must be occurring at substantial distances
($\leq$~36$\arcsec$, 1.3~kpc) above the galactic midplane.  This does
not include the even greater extent of Rand's shell 1 perpendicular to
the galactic midplane (z~$\lesssim$~1.7~kpc). Some unknown fraction of
the high diffuse $FUV$ emission at high altitude is likely due to dust
scattered light.

   Comparison of the $FUV$ and $H\alpha$  images indicates that the sizes 
of the \ion{H}{2} and star forming regions are about the same as well 
as being co-spatial. This implies that the B-type stars which 
account for most of the $FUV$ light have not moved far from their 
birth place. Typically, a star forming region which appears to be 
a single rounded area unblended with other regions in the UIT 
image subtends an angle of about 8~arcseconds corresponding to a 
linear distance of 291~pc at the assumed distance of 7.5~Mpc. 
Moving with a speed of 10~km~s$^{-1}$ stars would disperse to a width 
of 291~pc in 14~Myr. Although some of the $FUV$ bright regions are 
clearly blends of smaller regions making geometric comparisons 
impossible, we expect that many of the star forming regions must 
be on the order of 14~Myr old or younger.

\section{Photometry} 
                                     
\subsection{$FUV$ and Visual Photometry of 48 Regions} 
 
Figure~2 is the $FUV$ image which shows hand-drawn boundaries  and numbering of
the apertures in which photometry was performed.   The criteria for
establishing the numbered boundaries were (1)  that each should include at
least one unresolved $FUV$ emission  region that is also seen clearly in the B
image, (2) that each  should include all the $FUV$ reflection nebulosity
belonging to the  selected emission regions, and (3) that each should include
all  the compact \ion{H}{2} regions and $H\alpha$ filaments apparently
belonging to  the selected associations and not to the general background. The 
boundaries thus determined were also applied to the $U$, $B$, $V$, $H\alpha$ and
$H\beta$  images. The intent was to delineate those data which will  yield a
consistent model of star formation within each boundary.  Similarly, a large
aperture drawn around the entire galactic  image in each broad band and 
emission line permitted global  quantities to be computed from spatially 
integrated fluxes. 

The sky background, subtracted from each image, is determined from the
responses in 50 rectangular apertures distributed over the entire
field-of-view and located in clear regions between stellar images. The
apertures are 31$\times$31~pixels$^{2}$  and the standard deviation
of the distribution of the pixel responses within an aperture were
taken to indicate the rms variation in the pixel-to-pixel response. In
flux density units the pixel-to-pixel variation is
4.3~$\times$~10$^{-18}$~ergs~cm$^{-2}$~s$^{-1}$~\AA$^{-1}$ in the $FUV$
image and
$<$~1.4~$\times$~10$^{-18}$~ergs~cm$^{-2}$~s$^{-1}$~\AA$^{-1}$ in the
$U$, $B$ and $V$ images. The sky background is the average of the responses
in the 50 apertures. The standard deviation of the average responses
in the 50 apertures is taken to indicate the uniformity of the sky
over the field-of-view. This variation is
$<$~5.1$\times$10$^{-19}$~ergs cm$^{-2}$ s$^{-1}$~\AA$^{-1}$ in the
FUV image and
$<$~3.6$\times$10$^{-19}$~ergs~cm$^{-2}$~s$^{-1}$~\AA$^{-1}$ in the
$U$, $B$ and $V$ images. The variation in sky uniformity makes a negligible
contribution to the total random error in the derived flux
densities. However, the contribution of the pixel-to-pixel variation
can be as large as 28\%, although it is generally less than 10\%.

Correction for scattered light and sky background within the
individual apertures was achieved in the following manner. Each
aperture was expanded by 2~pixels (2.5~arcseconds) at its edge to create
annuli between the original and expanded apertures.  Histograms of the
pixel response to the background included in the annulus were made
from which the mean value was taken to be representative of the
background from all sources underlying the emission from the recently
formed stars and the gas excited by those stars. The total background
was simply the mean value multiplied by the number of pixels
encompassed by the original aperture.

Other than the correction for the sky background, no further
correction was made to the integrated flux density of each image
within the large aperture. Expanding and shrinking the large aperture
by 8~pixels or 20~arcseconds changes the integrated flux by 1\% or less
in all of the broad-band images. Thus, the integrated flux is quite
insensitive to the actual position of the large aperture on the sky.

The contribution to the observed $FUV$ continuum arising from  two photon
emission in the ionized gas was evaluated and found to  be generally at the 1\%
level with a maximum of 4\% of the observed  continuum. This effect was
neglected in subsequent calculations. 

Error will be made to the extent that the stellar emission, associated
reflected light and line emission from the individual regions
cannot be uniquely isolated from light emitted from the rest of the
galaxy. This is particularly true for the $B$, $V$, and $R$ images and is
exacerbated by the edge-on aspect of the object.  The problem is so
severe for the $R$-band image that the aperture photometry is very
inaccurate, and is for this reason not included in the subsequent
analysis.  In addition, uncertainties 
arising from the background correction procedure can be 
substantial, particularly for faint sources. Some indication of 
the magnitude of this systematic error can be had by expanding 
and contracting the apertures by 2~pixels (2.5~arcseconds). The 
result is that sky corrected flux densities change by an average 
factor of 1.5 for the broad-band photometry and factors of 1.4 and 
1.7 respectively for the $H\alpha$ and $H\beta$ line photometry. 

Listed in Table~2 are the photometry results corrected for background
and extinction in the Galaxy.  The latter correction was accomplished
using the Galactic value of E(B-V)~=~0.007 \citep{dev91} 
(see note b, Table~2) with
the \citet{sea79} extinction law.  The aperture number as in Figure~2
and CM \ion{H}{2} regions included in each aperture are listed in
column 1 and 2 respectively. The observed $FUV$ (152~nm), $U, B,$ and
$V$ fluxes in units of 10$^{-15}$~ergs~cm$^{-2}$~s$^{-1}$~\AA$^{-1}$
are listed in columns 3, 4, 5 and 6 respectively, and the $H\alpha$
and $H\beta$ fluxes in units of 10$^{-15}$~ergs~cm$^{-2}$~s$^{-1}$ are
listed in columns 7 and 8 respectively.  Values of $E(B-V)$ computed
from the Balmer Decrement are given in column 9.  The second number in
each column is the 1$\sigma$ standard deviation due to random errors
in the measurements only. The next to last row shows the sums of each
column for columns 2 through 7, i.e. the sum of the fluxes in all 48
apertures. The last row shows the flux values integrated over the
entire galactic image inside the large aperture.

%\begin{table}
%\dummytable\label{tab:2}
%\end{table} 

The $B-V$ color excesses were computed from Balmer Decrement 
and from the difference between the measured $B-V$
colors and $B-V$ colors determined from models of evolving star
forming clusters. The models (see Section~5, below) describe the
evolution of clusters in which the stars are all coeval, sometimes
called ``instantaneous burst'', or IB models. In principle, the derivation
of $E(B-V)$ in this way is more realistic than using the Balmer
Decrement since it is related to the extinction of stellar continuum
light rather than the extinction of emission from nearby excited
gas. However, the method suffers from the the need for 
{\it a priori} knowledge of the correct IB model parameters in order to 
infer $E(B-V)$. With the present data-set, the models were characterized
only by their burst ages which were determined from the ratio of the
number of Lyman continuum photons to the $FUV$ luminosity, and these
latter quantities were determined from the $FUV$ and $H\alpha$
observations corrected for dust extinction. Values for all other model
parameters were assumed as described below in Section~5.

   The distribution of $E(B-V)$ values determined from the Balmer 
Decrements is shown by the red line in Figure~3. A single 
gaussian fit to the distribution results in a mean value of $E(B-V) =$~0.39. 
Though the distribution is sharply peaked, there are 
broad wings at low occurrence levels extending to unrealistic 
negative $E(B-V)$ values. We take these broad wings to be the 
result principally of significant errors in the $H\beta$ flux density 
measurements. The less important part of the error is 
statistical; the most important part, as discussed above, is due 
to our inability to identify the relevant emission, and is 
estimated by the variable aperture photometry. Because of these 
errors we conclude that use of the individual color excesses to 
correct the measured flux densities for extinction is 
inappropriate. Rather, we determine a mean value to be applied to 
all the photometry by means of an iterative procedure. As an 
initial step, we use the mean value of the distribution computed 
from the Balmer Decrement to correct the $FUV$ and $H\alpha$ measurements 
in each region for extinction using the extinction law of 
\citet{cal94}. From these corrected values we compute models 
for each region which are characterized by unique burst times. 
Subsequently, $E(B-V)$ values are derived from the modeled and 
observed colors for each region from which a new mean value is 
determined. This process is repeated until the mean values of 
$E(B-V)$ converge. The final distribution of $E(B-V)$ values, shown 
as a black line in Fig 3, exhibits occurrence levels in the wings 
lower than those associated with the Balmer Decrement 
measurements, and a mean value of $E(B-V) =$~0.28 which is less by 
a factor of 0.7 than the value derived from the Balmer Decrement 
measurements. Qualitatively, this difference in derived values of 
$E(B-V)$ is characteristic of stars in regions of active star 
formation. However, the factor by which results of the two 
methods differ is usually smaller, being about 0.44 
\citep{cal94,fan88}.
The 2$\sigma$ width of the Gaussian function 
fitted to the final distribution is 0.36~mag. 

Several photometric quantities, which have been corrected for  internal
extinction, are listed in Table~3. As in Table~2, the  aperture number appears
in column 1. Magnitudes in the $FUV, U, B, $  and $V$ bands are listed in
columns 2, 3, 4 and 5 respectively. The  $H\alpha$ luminosity, the log of the
ratio of the number of Lyman  continuum quanta to the $FUV$ luminosity, 
log$_{10}$(N$_{Lyc}$/L$_{152}$), and the
M$_{152}$-$U$, M$_{152}$-$B$  and M$_{152}$-$V$ colors are listed in columns 6,
7, 8, 9 and 10  respectively. The associated formal random measurement errors
(1$\sigma$),excluding those arising from the extinction 
corrections are  listed adjacent to each quantity in each column.  
Uncertainty in the extinction correction will result in 
significant uncertainty in the listed quantities. Increasing or 
decreasing the mean value of $E(B-V)$ by the standard deviation in 
the $E(B-V)$ distribution changes the $FUV$, $U$, $B$ and $V$ magnitudes by 
0.80, 0.50, 0.43 and 0.36 mag respectively, the $H\alpha$ luminosity by 
a factor of 2 and the M$_{152}-U$, M$_{152}-B$ and M$_{152}-V$ colors by an average 
of 0.37~mag. The change in aperture sizes results in an average 
change in broad-band magnitudes of 0.56 mag, a change in the H$\alpha$
luminosity by 35\%, and an average change in the M$_{152}-U$, M$_{152}-B$ and 
M$_{152}-V$ colors of 0.37~mag. In general, random errors in quantities 
expressed as magnitudes are comparable to the changes induced by 
variation of the aperture sizes.  The errors do not include the 
systematic errors associated with the absolute calibration or the 
choice of the reddening law.

%\begin{table}
%\dummytable\label{tab:3}
%\end{table} 

\subsection{Integrated Photometry of NGC~4631} 
 
The last line in Table~2 gives the integrated photometry of that part
of NGC~4631 which can be observed, in particular all of the $FUV$ image inside
the large aperture. A comparison with two possible analog galaxies,
M101 (Sc(s)I) and M83 (SBc(s)II)), both observed face-on is given in
Table~4. Results from integrated UIT photometry of two additional
face-on galaxies, M74 (Sc(s)I) and M51 (Sbc(s)I-II), are also
listed. Excepting NGC~4631, the $FUV$ luminosities have been derived
from the results of \citet{wal97b},
with corrections only for extinction arising in the Galaxy, and with
modifications to take into account differences in the galactic
distances listed by \citet{wal97b} and
those used in this paper. The NGC~4631 152.1~nm flux is a factor of
1.1 brighter than the 155.0~nm flux
(9.81$\times$10$^{-13}$~ergs~cm$^{-2}$~s$^{-1}$~\AA$^{-1}$) measured
by \citet{cod82} using the University of
Wisconsin experiment on the first {\sl Orbiting Astronomical
Observatory} (OAO). The measured NGC~4631 $FUV$ luminosity is dimmer
by factors of 1.2 and 4.0 than the measured luminosities of M83 and
M101 respectively.

Far infrared (FIR) luminosities after
\citet{ric88}, adjusted for the differences in distances
between that publication and this paper, are listed in column 5. The
ratios of FIR to $FUV$ luminosity are listed in column 6. The
comparatively high L$_{FIR}$/L$_{FUV}$ ratio of NGC~4631 compared to
M101, M74 and M51 is expected because of its edge-on aspect. If, when
viewed face-on, the L$_{FIR}$/L$_{FUV}$ ratio for NGC~4631 were
similar to that of M101, i.e $\sim$4, then we can surmise that the
$FUV$ luminosity would be $\sim$2.3$\times$10$^{9}$~L$_{\sun}$ . If
the face-on L$_{FIR}$/L$_{FUV}$ ratio were more nearly like that of
M74 or M51, then the $FUV$ luminosity would be
$\sim$1.1$\times$10$^{9}$~L$_{\sun}$.  The exceptional galaxy in this
group of five is M83 which shows a comparatively large
L$_{FIR}$/L$_{FUV}$ ratio (nearly equal to that of NGC~4631) even
though it is viewed face-on. It may be that the dust in M83 provides
especially efficient conversion of $FUV$ light into FIR radiation
primarily because of its distribution relative to the hot, young
stars. The distribution of dust in M101 may also account for the
comparatively small L$_{FIR}$/L$_{FUV}$ ratio in that galaxy.
      
%\begin{table}
%\dummytable\label{tab:4}
%\end{table} 

\section{Cluster Ages, Masses, and Star Formation Rates} 
 
The photometric results can be combined with models of star forming clusters to
explore some aspects of the star formation processes in NGC~4631. The cluster
models we will use are discussed in detail in 
\citet{hil94}. Here, we will summarize that discussion. The models
are those of \citet{lan92}, which are based on the work of
\citet{leq81} who derive spectral energy
distributions and masses of evolving star clusters. The models use the 
Geneva stellar evolutionary tracks \citep{mey94} and the
models of \citet{kur91} at log$_{10}$(Z/Z$_{\sun}$)~=~-0.3. In
our work, the form of the IMF is $\Psi$(M) = dn(M)/d(lnM)~$\propto$~M$^{-x}$
for 1.8~M$_{\sun}$~$<$~M~$<$~120~M$_{\sun}$ and M$^{-0.6}$ for
0.007~M$_{\sun}$~$<$~M~$<$~1.8~M$_{\sun}$.  We assume an upper IMF slope of
x~=~1.35 conforming to the \citet{sal55} IMF. For our
purposes, an instantaneous burst (IB) model of a cluster of co-eval stars is a
spectral energy distribution (SED), L$_{IB}$($\lambda,t$), corresponding to a
given elapsed time from a burst of star formation. It is the sum of the SEDs of
all the stars in the cluster at age $t$ weighted by the assumed IMF and
normalized to an astrated mass of 1~M$_{\sun}$.  Its units are 
ergs~s$^{-1}$~\AA$^{-1}$~M$_{\sun}$$^{-1}$. The ionizing photon luminosity is
computed by integrating L$_{IB}$($\lambda,t$)/hc over $\lambda$ shortward of
912~\AA.  The observed ionizing photon luminosity is given by

\begin{equation} 
              N_{Lyc} = 2.206~L_{H\alpha,0}\lambda_{H\alpha}/hc 
\end{equation} 
\noindent
where L$_{H\alpha,0}$ is the $H\alpha$ luminosity obtained from the
observed flux corrected for both extinction and distance. The factor
2.206 is derived under the assumption of case B recombination at
10,000K.  The $FUV$ luminosity, L$_{IB}$(1520~\AA,$t$), is defined as
L$_{IB}$($\lambda,t$) integrated over the $FUV$ response curve of the
UIT. The age of a cluster is simply determined by finding that value
of $t$ at which the observed and model ratios of
N$_{Lyc}$/L$_{IB}$(1520~\AA,$t$) are equal.  The total astrated mass
inferred from the IB model is simply the observed luminosity divided
by the normalized IB model SED integrated over the response curve for
each broad band.
 
If continuous star formation (CSF)has occurred at an essentially
constant rate over a period of time, the star formation rate, SFR, can
be computed by integrating the IB models over time; i.e.

\begin{eqnarray}
SFR_{\lambda} = {{L_{obs,\lambda}}\over{\int_{t} L_{IB}(\lambda,t)dt }}~ M_{\sun} yr^{-1}
\end{eqnarray}
\noindent 
where L$_{obs,\lambda}$ is the measured spectral luminosity in
ergs~s$^{-1}$~\AA$^{-1}$ and L$_{IB}$($\lambda,t$) is the IB model defined
above. The integration in time is over the interval of star formation.
 
\subsection{Instantaneous Burst Models} 
 
Plots of the observed colors, M$_{152}$-$U$, M$_{152}$-$B$, and
M$_{152}$-$V$, (corrected for extinction),
vs. log$_{10}$(N$_{Lyc}$/L$_{152}$) are shown in Figure~4, a, b, and
c. Numbers adjacent to the data points in Figure~4c identify the
aperture corresponding to each datum. Error bars indicate the
1$\sigma$ random photometric errors.  The points which have been
plotted meet the conditions that the signal to noise ratios in the
M$_{152}-V$ colors are greater than 3 and the color changes due to
variations in the aperture sizes are less than the 1$\sigma$ errors.

The triangle symbols represent the colors computed from IB models
where each triangle corresponds to a time in the past when the star
burst occurred in integer units of 1~Myr beginning at 2~Myrs on the
right and ending at 11~Myrs on the left. The arrow indicates the
direction points would move if $E(B-V)$ were to increase by 0.18~mag
which is the 1$\sigma$ width of the relevant $E(B-V)$ distribution
(black line) shown in Figure~3. All three colors derived from the
observations of the selected star forming regions tend to be more
positive than those derived from the models. This behavior may be due
to our use of an inappropriate reddening curve or possibly to the
assumptions regarding the model input parameters. However, in view of
the random and systematic errors such distinctions are
problematical. For example, increasing the mean value of $E(B-V)$ by
about 0.11~mag, which is the adopted uncertainty in this quantity for
integrated values as explained below, brings the data in reasonably
good agreement with the models.  A comparison of the observational
results with the IB model indicates that if the star formation occurs
in bursts then the bursts occurred from 2.7 to 10~Myrs in the past. It
will be noted that the burst times are strongly dependent on
log$_{10}$(N$_{Lyc}$/L$_{152}$).  On average, the change in aperture
size will move the points to the left or right an amount equivalent to
1~Myr. The implication of the uncertainty in the extinction correction
is that star formation for some regions may have occurred 2.7~Myr
further in the past, or 2.3~Myr toward the present.

Individual burst times, determined from the measured
log$_{10}$(N$_{Lyc}$/L$_{152}$) values are computed for each
aperture. The average variation in burst times attributable to the
photometric standard deviation is 0.71~Myr, which is about the same as
variation due to changes in the aperture sizes. The total astrated
mass is determined for all apertures from photometry in the four wide
bands as outlined above. The masses summed over the 48 apertures for
each broad band are similar; the largest deviation, 20\%, is
associated with the $FUV$ photometry. This result is remarkable in
that the mass of the stars producing the $FUV$ light constitutes a
small part of the total mass of stars producing light in the optical
bands, and it implies that an average of the summed masses in each
band is justified. Random errors in the masses summed over all the
apertures for each band are on the order of a percent whereas changes
in the aperture sizes cause the summed masses to vary by a factor of
1.41. For integrated quantities corrected for extinction internal to
NGC~4631 we adopt, somewhat arbitrarily, an uncertainty in $E(B-V)$ of
0.11, which is equal to the difference between the mean values of the
$E(B-V)$ distributions derived from the Balmer Decrements and from the
stellar continua. This difference results in an uncertainty in the
summed masses by of factor of 2.8. When the aperture size variation is
combined with the extinction uncertainty the summed burst mass is
uncertain by a factor of 3.9.
 
\subsection{Continuous Star Formation Models} 
 
\citet{cal97} has suggested that in active star forming regions such
as described here, B type stars, which supply large amounts of $FUV$
radiation but little $H\alpha$, migrate away from the dust clouds of
their origin where star formation is continuing to produce massive O
stars. A typical migration time of 50~Myr corresponds to a linear
dimension of 500~pc and a stellar velocity of 10~km~s$^{-1}$. Such a
picture can explain why the color excesses measured from stellar
continuum colors are often less than those measured from Balmer
Decrements in the gas surrounding Lyman continuum emitting stars.
From our observations we estimate that the smallest, most compact
UV-bright regions in NGC~4631 can not have begun forming stars
substantially more than 14~Myr ago purely on the basis of the
morphological similarity of the H$\alpha$ and FUV images.  The
crossing time of the largest region is equivalent to a turn-on time of
48~Myr in the past.  However, the FUV image must be composed in part
of many spatially unresolved clusters, and with the available data we
can not disentangle effects due to the diffusion of stars from those
due to the distribution of stellar birth sites.  The impression of
comparative youth is enhanced by the similarity of the mean E(B-V)
values determine by the Balmer Decrements and the integrated stellar
continua of each star forming region.  Nevertheless, there is likely
to be a spread in turn-on times among the UV-bright regions.  The analysis
presented in Section 6 implies that the large eastern shell may have
resulted from a star burst which ocurred 20~Myr ago.  Somewhat
arbitrarily, we assume that the most representative turn-on time is
14~Myr in the past, but that this representative turn-on time could
vary by at least $\pm$~5~Myr.

Shown in Figure~4 a, b, and c are colors determined from CSF models
with star formation beginning 14~Myrs ago and ending at times
indicated by circles plotted in 1~Myr intervals from the present
(0~Myrs) on the right to 9~Myrs ago on the left.  The data points
indicate a similar range in turn-off times. Star formation turn-off
times and astration rates computed from relation (2) for each passband
and the $H\alpha$ line are listed in Table~5, columns 8 through
13. The rate averaged over all four passbands and the $H\alpha$ is
listed in column~14. Omitted from this compilation are the rates for
region 28, which shows a turn-off time so close to the onset of star
formation that the rate is unrealistically high. Stated differently,
region 28 must be considered a star burst.  The average star SFR
summed over all the star formation regions is
0.33~M$_{\sun}$~yr$^{-1}$.  If the representative turn-on time was
20~Myr ago, then the total SFR would be about half this rate, or
0.15~M$_{\sun}$~yr$^{-1}$.  The average variation of the summed SFR due
to the change in the aperture sizes is $\pm$0.17
M$_{\sun}$~yr$^{-1}$, whereas error in the extinction correction (0.11~mag)
results in an uncertainty in the summed SFR of a factor of 5.5. The
random error in the summed average SFR is about 1\%.

%\begin{table}
%\dummytable\label{tab:5}
%\end{table} 

The diagnostic diagrams derived for the CSF models are very similar to
those derived for the IB models, their differences being significantly
smaller than the dispersion of the data. The observed colors agree
with the model colors within the limits defined by the estimated
errors. The slope displayed by the CSF models is flatter and less
variable than that of the IB models, particularly at the most recent
turn-off times (or burst times).  This circumstance arises from the
fact that most of the FUV flux is provided by B-type stars with
lifetimes longer than 14~Myrs.  Therefore, the relative proportion of
B and later type stars does not vary significantly over the 14~Myr
baseline, thus yielding approximately constant UV colors which vary
smoothly with turn- off time.

\section{The Eastern Supershell} 
 
\citet{mcc87} have developed a
quantitative picture of the dynamics of supershells, which can be used
to interpret our results and reinterpret those of 
\citet{ran93}.  The picture is one in which the
supershell was formed and driven to its present size primarily by
supernovae occurring in an extraordinarily large collection of
associations and clusters resembling those we observe. Combining
equations (3) and (4) of \citet{mcc87} gives an expression

\begin{equation}
       t_{7} = 0.0588(km~s^{-1}~pc^{-1})~(R_{s}/V_{s}) 
\end{equation}
\noindent
where t$_{7}$, in units of 10~Myr, is the time required for the shell
to expand to a radius, R$_{s}$, with a terminal velocity of V$_{s}$.
Adopting R$_{s}$ = 1.5~kpc and V$_{s}$~=~45~km~s$^{-1}$ 
\citep{ran93} we find t$_{7}$~=~1.96 or about 20~Myr.  During this
interval, only those stars with mass equal to or greater than M$_{*}$
will become supernovae and contribute to the shell expansion. M$_{*}$
is given approximately by the relation
M$_{*}$~$\sim$10~M$_{\sun}$~(t$_{7}$/3)$^{-0.625}$ 
\citep{mcc87}, and is 13~M$_{\sun}$ for the assumed
conditions.  The total number of stars, N$_{*}$, contributing to shell
expansion is given by

\begin{equation}
         N_{*}~\times~E_{51} = n_{0}~(R_{s}/82 pc)^{5}~t_{7}^{-3}
\end{equation}
\noindent
where E$_{51}$ is the energy input to the ISM from each supernovae
event in units of 10$^{51}$~ergs, here assumed to be unity, and
n$_{0}$ is the density of the ambient ISM, which is assumed to be
uniform.  Substituting the numerical values for R$_{s}$ and t$_{7}$
gives N$_{*}$~=~2.7$\times$10$^{5}$~n$_{0}$ stars with masses
$\geq$~13~M$_{\sun}$.  If n$_{0}$~=~0.2~cm$^{-3}$, consistent with the
estimate of \citet{ran93}, then
N$_{*}$~=~5.3$\times$10$^{4}$ stars (M $\geq$ 13 M$_{\sun}$). This
number is somewhat larger than the estimate of 
\citet{ran93} (1 - 3.5)$\times$10$^{4}$ stars, and in
absolute terms is extraordinarily large. For example, \citet{ken84} 
has estimated that in 30~Doradus the total mass
of OB stars between 10 and 100~M$_{\sun}$ is 50,000~M$_{\sun}$.
Assuming a \citet{sal55} IMF, the total number of
OB stars between 13~M$_{\sun}$ and 100~M$_{\sun}$ is
1.54$\times$10$^{3}$, a factor of about 34 less than our estimated
number of OB stars responsible for the large shell. We find that the
light emitted from region 20, which includes the giant \ion{H}{2}
region CM~67, is produced by stars with a total mass of
2.7$\times$10$^{6}$~M$_{\sun}$.  If our assumed IMF is valid, then
this region contains approximately 1.3$\times$10$^{4}$ stars between
13~M$_{\sun}$ and 120~M$_{\sun}$, a number which approaches that
inferred for the ancestral stars of the large shell. We infer an even
larger number of stars (1.5$\times$10$^{4}$) for region 28.

In this picture of shell formation, the UV-bright regions associated
with the shell, namely those delineated by apertures 3 through 16,
must arise from second generation star formation. As 
\citet{mcc87} point out, the shell will probably
burst at high galactic latitudes before third generation stars are
formed. In fact, as \citet{ran93}
suggest on the basis of a break in the \ion{H}{1} emission observed on
the north side of the shell (their Figure~3), this breakout may have
already occurred. These authors also point out that a shell of 3~kpc
diameter can be produced initially only if there is significant gas
pressure and/or magnetic pressure at high z.  These conditions seem
plausible when taking into account the obvious tidal effects of the
interaction between NGC~4631 and NGC~4656 manifested in the \ion{H}{1}
emission maps of \citet{wel78} and \citet{ran93}.  
 
If the stars in the vicinity of the supershell were made from a reservoir of
gas, most of which was swept up into the shell, then it is possible to make an
estimate of a star formation efficiency, here defined as
M$_{stellar}$/(M$_{stellar}$ + M$_{gas}$), in the supershell region by
referring to the \citet{ran93} estimate of the hydrogen supershell
mass. The average total stellar mass in the shell region (apertures 3 through
16) inferred from our observations in the $FUV$, $U$, $B$, and $V$ bands is
1.6$\times$10$^{7}$~M$_{\sun}$ based on the IB model.  Comparison with the
estimated hydrogen shell mass of (1-2)$\times$10$^{8}$~M$_{\sun}$ implies an
efficiency of $\sim$0.047. This efficiency represents an upper limit since the
contribution of a molecular gas has not been taken into account.

\section{Discussion} 
  
Interpreted in terms of an instantaneous burst model, the  light in all the
numbered apertures and in all four wavelength  bands implies a total created
mass of 2.5$\times$10$^{7}$~M$_{\sun}$ within a factor of 3.9.  
This number  may be underestimated by a comparatively small amount due
to the absorption of light by  interstellar grains as discussed below. Our
estimate can be  compared to the equivalent number of
5.2$\times$10$^{6}$~M$_{\sun}$ in the  Magellanic irregular, NGC~4449, which
exhibits intense star  formation \citep{hil94}. The
difference between the two  galaxies is even greater than indicated here
because NGC~4449 is  observed nearly face-on revealing more of its
star-forming  regions than does NGC~4631 which is observed almost edge-on. 

The interior disk of NGC~1068 exhibits numerous starburst knots surrounding the
nucleus. \citet{bru91} have estimated the total
mass of nine of the brightest of these knots, assuming an IB scenario, to be
2.4$\times$10$^{7}$~M$_{\sun}$, a number comparable to that derived for our observed 
48 regions in NGC~4631.  Thus, while the masses deduced for the
star-forming regions of NGC~4631 are extraordinarily large, they are not
uniquely so.
 
We also compare the star formation activity of the individual star
forming regions with the prototypical giant \ion{H}{2} region,
30~Doradus.  Using an aperture of 370~pc (d=50~kpc), 
\citet{ken84}, obtained L$_{H\alpha}~=~$1.5$\times$10$^{40}$~ergs~s$^{-1}$ 
and an estimated total mass of 50,000~M$_{\sun}$ for all stars with
individual masses between 10 and 100~M$_{\sun}$ .  Extending this
Kennicutt estimate over a larger mass range by using the Salpeter IMF
we find a total stellar mass between 0.007 and 120~M$_{\sun}$ of
1.11$\times$10$^{6}$~M$_{\sun}$.  If it is assumed that this mass was
produced in a burst, we can compare the total mass estimate with the
IB model mass estimates for the individual star forming regions listed
in columns 3, 4, 5 and 6 of Table~5. Of the 48 aperture delineated
\ion{H}{2} regions listed, 6 exhibit derived total masses, averaged
between the four wide-band images, in excess of the 30~Doradus value.
Taking into account our estimated error, this number can be increased
to 26. Of these 6 regions, none exhibit $H\alpha$ luminosities comparable to
30~Doradus. Again, the photometric uncertainty allows this number to be
increased to 3.  The $H\alpha$ luminosity of region 20 is about two-thirds that
of 30~Doradus, but its mass is about a factor of 2.5 greater. Region 20
incorporates the bright \ion{H}{2} region CM~67 and is associated with a
giant molecular cloud, both of which have been tentatively identified
with the tip of a bar \citep{roy91}.  Regions 41 and 47
show both $H\alpha$ luminosities and masses comparable to 30~Doradus, and we note
also that the $H\alpha$ luminosity of region 41 is comparable to that of the
supergiant \ion{H}{2} region in M101 (NGC~5461), which has an $H\alpha$ luminosity
uncorrected for extinction of 2.7$\times$10$^{40}$~ergs~s$^{-1}$ \citep{wal90}. 
However, with some significant exceptions such as those
highlighted above, most of the NGC~4631 star forming regions are less
massive and older than 30~Doradus. Even when the upper limit on $E(B-V)$ is
used to correct for internal extinction only 7 regions exhibit $H\alpha$ 
luminosities about equal to or exceeding that of 30~Doradus.

A ``short term'' star formation rate (SFR) averaged over the range in
burst times exhibited by the data, i.e. 7.3~Myrs, can be computed using
the created masses summed over all 48 apertures and averaged over the
four broad bands. This rate is $\sim$3.5 M$_{\sun}$~yr$^{-1}$ which is
substantial.  As discussed below, it is about equal to the global SFR determined 
from FIR measurements, and a factor of 7 less than the
total SFR measured by \citet{dah95}.

In summary, when the $FUV$ and optical imagery is interpreted in  terms of the
instantaneous burst paradigm it is obvious that NGC~4631  incorporates
exceptionally large star formation activity. 

Interpretation of our results in terms of CSF models points to a
galaxy of a more ordinary variety.  The total average star formation
rate determined from the $FUV$ data summed over the 48 apertures for
NGC~4631 is 0.33~M$_{\sun}$~yr$^{-1}$, which is a factor of 2.7 less
than than the equivalent NGC~4449 rate (0.88~M$_{\sun}$~yr$^{-1}$).
However, our assumed uncertainty in the mean value of $E(B-V)$ allows
our derived SFR to exceed the NGC~4449 rate by a factor of 2. Thus,
that part of NGC~4631 which is observable and has incorporated star
formation within the past 14~Myr creates stars at a rate comparable to
the total rate of the face-on starburst galaxy, NGC~4449.  Assuming
that the turn-on time was 20~Myr ago, which
results in an SFR equal to 0.15~M$_{\sun}$~yr$^{-1}$, does not change
this conclusion.  The NGC~4631 results can also be compared to the
high rate of 2.4~M$_{\sun}$~yr$^{-1}$ found by
\citet{bru91} when interpreting in terms of a CSF scenario 
their observations of the star forming regions found in the disk of
NGC~1068.  Assuming constant star formation, our observations indicate
a total mass created in 14~Myr of about 4.6$\times$10$^{6}$
M$_{\sun}$, which is less than the mass inferred from the IB model
(5.1$\times$10$^{7}$ M$_{\sun}$) by a factor of 5.5.  This factor
increases to 8.4 if the turn-on time was 20~Myr ago.

\citet{dah95} have related radio continuum
observations  of the halos of several edge-on galaxies, including NGC~4631, to 
the rates of supernovae occurring within the galactic disks.  The 
inferred supernova rate for NGC~4631 is 0.23~yr$^{-1}$. Assuming
equilibrium between the birth rate and death rates of  supernovae producing
stars and a \citet{sal55} IMF over the mass range 
between 0.007 and 120~M$_{\sun}$ we find that this result implies a star 
formation rate of 24~M$_{\sun}$~yr$^{-1}$.  

A significantly smaller rate is derived from the FIR luminosity.  
For stars with masses greater than 10~M$_{\sun}$ 

\begin{equation}
             SFR_{FIR} (M_{\sun}~yr^{-1}) = 1.4 \times 10^{-10} L_{FIR} (L_{\sun})
\end{equation}
\noindent
\citep{sau92}.  In the case of NGC~4631,
SFR$_{FIR}$(M$_{\sun}$ $>$ 10) = 1.3~M$_{\sun}$~yr$^{-1}$.  Employing
the Salpeter IMF described in Section~5, the SFR for stellar masses
between 0.007~M$_{\sun}$ and 120~M$_{\sun}$ is
8.1~M$_{\sun}$~yr$^{-1}$ or 4.5~M$_{\sun}$~yr$^{-1}$ if 45\% of the
FIR luminosity is from the diffuse interstellar medium.

The total continuous SFR, summed over the 48 apertures in our data
set, 0.33~M$_{\sun}$~yr$^{-1}$, is far smaller than the
rate inferred from Dahlem's radio observations. It, of course,
refers only to the star forming activity delineated by the 48
apertures.  If it is assumed that the recorded $FUV$ emission
from the whole galaxy arises from star formation, then multiplication of the
0.33~M$_{\sun}$~yr$^{-1}$ rate by the ratio of the observed $FUV$
flux in the large aperture to the observed flux summed over all 48
small apertures increases the continuous SFR to
2.2~M$_{\sun}$~yr$^{-1}$, still a very small SFR compared to the Dahlem value 
but one that approaches the FIR rate.  

It is possible that a large portion of the newly created stars are so
heavily shielded by dust that there is no indication of their
existence in any of our wavelength bands. For example, 
\citet{bru91} estimate that the emerging UV
light from $FUV$ bright knots in the disk of NGC~1068 implies an SFR
which is only 7\% of the true value. This estimation was based on the
relative FIR and UV knot luminosities and the assumption that most of
the FIR emission is reprocessed UV radiation from hot stars absorbed
and reradiated by interstellar grains. Unfortunately, no far infrared
observations of NGC~4631 exist at a spatial scale which would allow
the isolation of FIR emission from the galactic background for any of
the star formation regions. An alternative procedure for approximating
the relative contribution of reprocessed UV radiation in NGC~4631
star-forming regions is to measure the ratio of integrated $FUV$ and
FIR luminosities of face-on galaxies. The galaxy M74 (NGC~628) is
ideal for this purpose since the required luminosities are known and
$E(B-V) = $0.33 over much of the galaxy 
\citep{cor94}, a value which is  similar to the mean value of the color excesses
measured for the NGC~4631 star-forming regions. The ratio of the total
integrated FIR luminosity from IRAS measurements 
\citep{ric88} to the integrated $FUV$  luminosity from 
UIT measurements \citep{wal97b} is 8.5
(see Table~4). Similar ratios for other face-on galaxies are also
listed in Table~4.

It is likely that the dust emitting the $IR$ radiation is heated in part by 
the interstellar radiation field, with an approximate contribution to the 
total $IR$ heating budget of 45\% \citep{sau92}; 
though see contrary arguments of \citet{dev92}
and \citet{dev94} who claim that almost all of 
the $IR$ radiation is reprocessed $FUV$ emission.  Assuming that 
such processes are occurring in NGC~4631, and adopting the M74 
$FIR$/$FUV$ luminosity ratio of 8.5 (the average $FIR$/$FUV$ luminosity 
ratio for the face-on spiral galaxies M83, M101, M74, and M51 is 8.8) 
and an interstellar to total $IR$ radiation energy density 
ratio of 0.45, we find that the SFR determined from the 
{\it extinction corrected FUV} luminosity will be increased by a factor 
of only 1.21 when taking into account 
the stellar $FUV$ light absorbed by the interstellar grains.  
Thus, the total SFR for the 48 
star forming regions alone becomes 0.40~M$_{\sun}$~yr$^{-1}$ with a large 
uncertainty. If the 
integrated $FUV$ emission within the large aperture is assumed to arise 
entirely from star formation then the total estimated SFR {\it in that part 
of NGC~4631 which we are able to observe} becomes 2.7~M$_{\sun}$ yr$^{-1}$.

   The correction is not sensitive to assumptions about the IR 
heating budget. For example, if all the observed FIR light was 
reprocessed $FUV$ light from star formation instead of only 55\% as 
we have assumed then the correction factor would increase by only 
16\%. On the basis of observations of M83 and M101 the uncertainty 
in the value of R, i.e. the applicable value of the ratio of 
observed FIR to $FUV$ light, induces an uncertainty in the 
correction factor of 12\%. However, the correction is still quite 
uncertain since FIR observations of NGC~4631 with spatial 
resolution comparable to those of the UIT have not been made.

If all the FIR light is reprocessed $FUV$ light, then the SFR inferred
from the $FUV$ light in the large aperture becomes
3.1~M$_{\sun}$~yr$^{-1}$, again with large uncertainties, which should
be compared to the SFR computed from the observed FIR emission via
relation (5) of 8.1~M$_{\sun}$~yr$^{-1}$. The result that the SFR
summed over the 48 apertures is from factors of 11 to 25 less than the
SFR determined from the FIR observations is not surprising considering
the edge-on aspect of NGC~4631. Although the amount of created stellar
mass differs by a factor of 5.5 there is no reason for distinguishing
between the applicability of the IB and CSF models on the basis of the
diagnostic diagrams in Figure~4.  The fact that the large aperture
$FUV$ photometry leads to an SFR about equal to that derived from the
FIR within a factor of 2 is surprising, and may indicate that most of
the FUV light arises from short term star bursting activity, i.e. that
the exclusive application of the CSF model is not appropriate.

By contrast, the SFR inferred from the radio halo 
\citep{dah95} is a factor of at least 5 larger than that
associated with either the large aperture FUV photometry or the FIR
results (assuming only 55\% of the observed FIR is reprocessed FUV
light).  We speculate that the SFR derived from the radio observations
is too large because the assumption of equilibrium between the birth
and death rates of the supernovae producing stars is invalid, and that
the radio halo is due to a starburst and not continuous star
formation.

Finally, to the extent that the  IB paradigm is valid, our
conclusions regarding the  formation of the supershell in NGC~4631 indicates
that strong  starburst activity has occurred in that part of NGC~4631 
approximately 20~Myrs ago.
 
\section{Summary} 
 
A far ultraviolet image of the galaxy NGC~4631 centered on  152.1~nm was
recorded by the Ultraviolet Imaging Telescope. The  $FUV$ image, combined with
similar images in the $U, B, V,$ and $R$  bands and in $H\alpha$ and $H\beta$,
is used to study star formation in this  late type (Sc/SBd) galaxy, viewed
nearly edge-on, with the  following results.  
 
1. For a disk galaxy seen edge-on, NGC~4631 is extraordinarily  bright at $FUV$
wavelengths. The attenuation of light from 48 star  forming regions located
uniformly along the whole length of the  visible galactic disk is low with the
mean value of $E(B-V) = $0.28.   Dust is seen silhouetted against the cool star
population,  and the star forming regions appearing in $H\alpha$ emission are
seen  silhouetted against the dust. These facts together with our 
interpretation of previously measured \ion{H}{1} and \ion{H}{2} 
position-velocity curves indicate that the $FUV$ emission arises 
near the  edge of the observable galaxy, as is observed in the 
face-on Sc/SBc galaxies M101 and M83. 

The overall $FUV$ morphology of this edge-on disk galaxy is  remarkably similar
to that of the so-called chain galaxies that  have been recently observed  at
high redshift, thus implying a  similar interpretation for at least some of
those distant $FUV$  sources (see \citet{smi97}). 

2. Photometric data is presented in the form of diagnostic 
diagrams showing the M$_{152}-U$, M$_{152}-B$ and M$_{152}-V$ colors plotted 
against log$_{10}$(N$_{Lyc}$/L$_{152}$). The data exhibit considerable dispersion 
due to errors made in the determination of the relevant 
background and extinction corrections. Both instantaneous burst 
(IB) models with burst ages ranging from 2.7 to 10~Myr and 
continuous star formation (CSF) models with star formation 
beginning 14~Myr ago and terminating from 0 to 9~Myr ago fit the 
data equally well. The uncertainties permit the burst ages and 
turn-off times to be extended 1 to 3 Myrs to earlier times. 

3. The ratio of the mean value of $E(B-V)$ determined from the 
continua of stars to that found from measurements of Balmer 
Decrements is 0.7, and is thus greater than the value of 0.44 
quoted by \citet{cal97} as being typical of regions exhibiting 
active star formation. This result is compatible with our 
conclusion on the basis of $FUV$ emission morphology that the 
observed star formation is young, i.e. that the stars have not 
had much time to migrate from their birth site.  
    
4. If the observational results are interpreted with the IB models,
the inferred stellar masses created in each star forming region are
substantial. As determined by our analysis, the inferred stellar
masses of 6 regions exceed the mass of 30~Doradus.  Errors permit this
number to increase or decrease by up to 20.  The region of the second
largest derived mass, 2.7$\times$10$^{6}$~M$_{\sun}$ , incorporates
the bright \ion{H}{2} region CM~67 and is associated with a giant
molecular cloud, both of which have been tentatively identified with
the tip of a bar \citep{roy91}. Of these 6
regions, none exhibit $H\alpha$ luminosities equal to that of
30~Doradus. Photometric uncertainties permit this number to increase
to 3. The $H\alpha$ luminosity of region 41 is comparable to that of
the supergiant H II region in M101 (NGC 5461).  The total created
stellar mass deduced from the light emitted from the star forming
regions is 2.6$\times$10$^{7}$~M$_{\sun}$ to within a factor of 2.3,
which is about equal to the stellar mass
(2.4$\times$10$^{7}$~M$_{\sun}$) of the UV emitting knots in the disk
of NGC~1068 \citep{bru91} and is a factor of 5 greater than the stellar
mass (5.2$\times$10$^{6}$~M$_{\sun}$) associated with the $FUV$
emitting knots in NGC~4449 \citep{hil94}.
Using the range in burst times of 7.3 Myrs, we derive an average
``short term'' star formation rate of 3.5~M$_{\sun}$~yr$^{-1}$.

5. Based on previously obtained data, application of the shell
generation theory of \citet{mcc87} shows
that the number of supernova producing stars which could generate the
3~kpc diameter ring of $FUV$ and \ion{H}{1} emission near the east end
of the galactic image is 5.3$\times$10$^{4}$, which is somewhat
greater than the range (1 to 3.5)$\times$10$^{4}$ estimated by 
\citet{ran93} based on the dynamics of a
hydrogen supershell surrounding the stellar ring. While this number is
extraordinarily large, it is comparable to the same number,
1.3$\times$10$^{4}$, associated with the intense star forming region
delineated by aperture 20, a region that includes the giant \ion{H}{2}
region CM67.  Those FUV bright regions on and within the periphery of
the large shell (apertures 3 through 16) must represent second
generation star formation, and this conclusion together with the shell
analysis implies that strong star formation occurred approximately
20~Myrs ago in that part of NGC~4631 incorporating the large shell.

6. Analysis of the $FUV$, $U$, $B$, and $V$ photometry in the context
of the CSF paradigm yields a total SFR for the 48 individual star
forming regions equal to 0.33~M$_{\sun}~$~yr$^{-1}$, uncertain by a
factor of 8.2.  Approximate accounting for UV light which is absorbed
by interstellar grains leads to a corrected rate of
0.40~M$_{\sun}~$~yr$^{-1}$.  Under the assumption that 55\% of the
observed FIR is from reprocessed $FUV$ light this corrected rate is
about a factor of 11 less than the SFR inferred from the measured
total $FIR$ (4.5~M$_{\sun}$~yr$^{-1}$), which is not surprising
considering the edge-on aspect of NGC~4631. If the observed $FUV$
light integrated over that part of the galaxy we are able to observe
is assumed to arise in constant star formation, then the SFR equals
2.6~M$_{\sun}$~yr$^{-1}$ when approximate accounting for the
reprocessing of $FUV$ light by interstellar grains is taken into
account. This is a surprising result in that one would expect the SFR
inferred from the integrated $FUV$ data to be significantly less than
the global SFR determined from the FIR observations, and the
photometric uncertainties exacerbate the problem. We interpret this
paradox to indicate that the $FUV$ emission should not be seen
exclusively in terms of the CSF model, but that much of the observed
global $FUV$ light arises in star bursts. This interpretation is
consistent with the inference of star bursting activity near the
galactic center drawn by \citet{gol94b}
and in an inner disc (r$\approx$50$\arcsec$) by \citet{dur82}.

\acknowledgements

We thank the many people who made the ASTRO 2 mission a success. We would also
like to thank Robert Hill for his advice on many aspects of image data
reduction and Richard Rand for providing us with a calibrated $H\alpha$ image
of NGC~4631. Funding for the UIT project was through the Spacelab Office at
NASA Headquarters under project 440-51. This research has made use of the
digitized version of the National Geographic Palomar Sky Survey produced at the
Space Telescope Science Institute under U.S. Government Grant NAGW-2166. The
Oschin Schmidt Telescope is operated by the California Institute of Technology
and Palomar Observatory. W.W.W. acknowledges partial support from NASA through
the Astrophysical Data program (071-969dp).

\clearpage

\begin{figure}
\figurenum{1}
\begin{minipage}{6.25in}
\plotone{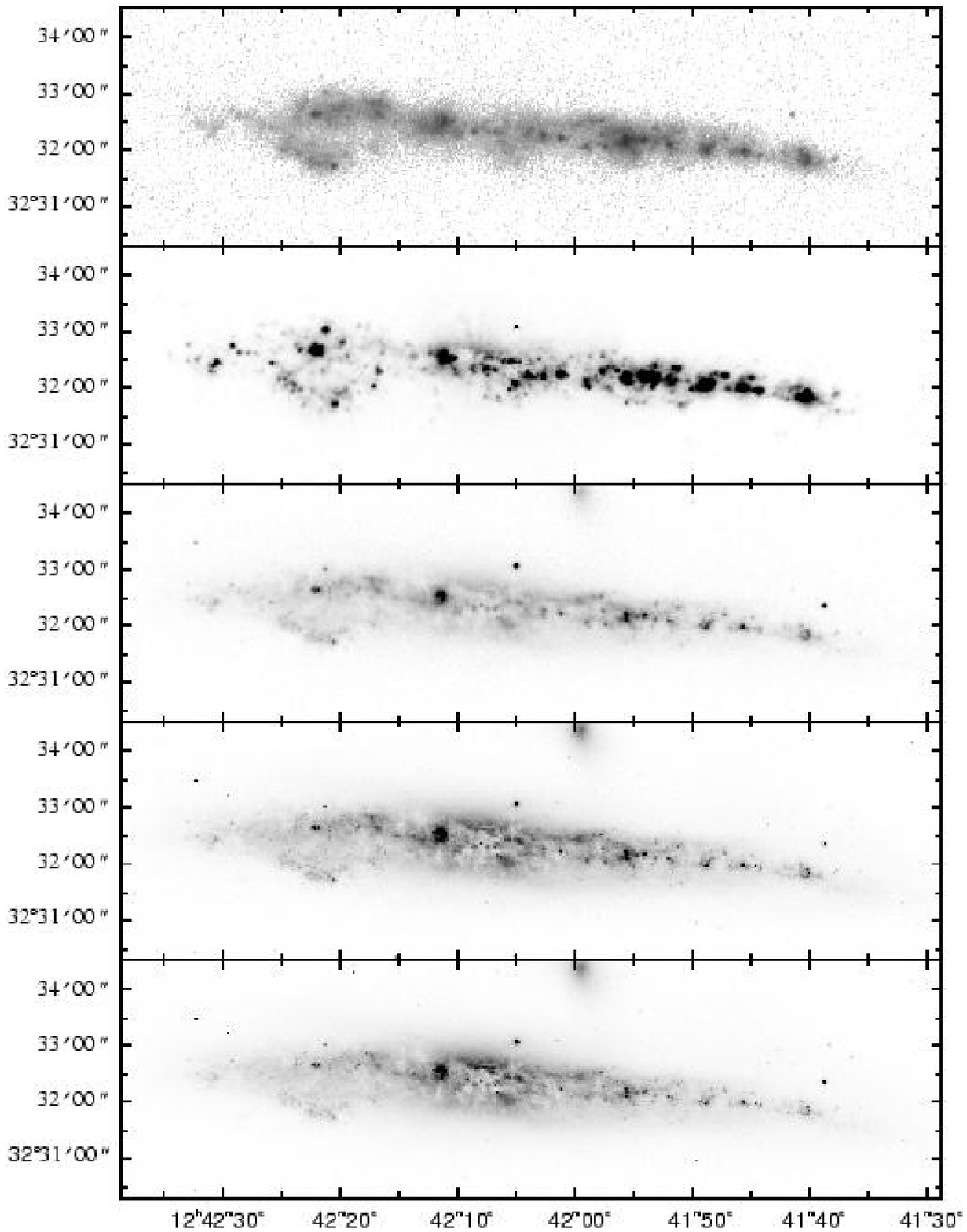}
%\plotone{f1.orig.eps}
\caption{Far-ultraviolet (FUV), $H\alpha$, 
$U$, $B$, and $V$, images of NGC~4631. North is up, east to the left. 
Portions of the dwarf elliptical galaxy, NGC~4627, which is part of
the interacting system composed of NGC~4631, NGC~4656 and NGC~4627,
can be seen on the north side of the $U$, $B$, and $V$
images.}
\label{fig1}
\end{minipage}
\end{figure}

\clearpage

\begin{figure}
\figurenum{2}
\begin{minipage}{6.25in}
\plotone{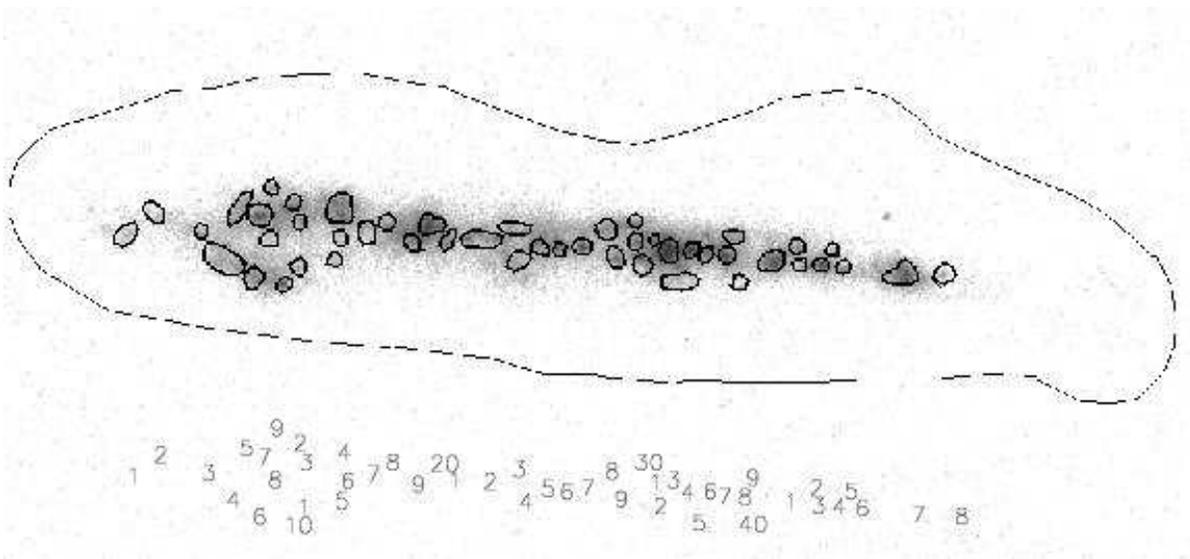}
%\plotone{f2.orig.eps}
\caption{$FUV$ image of NGC~4631 as in Figure~1 with aperture
boundaries indicated and numbered. The apertures are numbered, 1 to
48, from left to right, but because of crowding only the units number
is shown except for apertures numbered 10, 20, 30 and 40.  The large
aperture boundary encompassing the whole image is used to compute
global properties.}
\label{fig2}
\end{minipage}
\end{figure}

\clearpage

\begin{figure}
\figurenum{3}
\begin{minipage}{6.25in}
\plotone{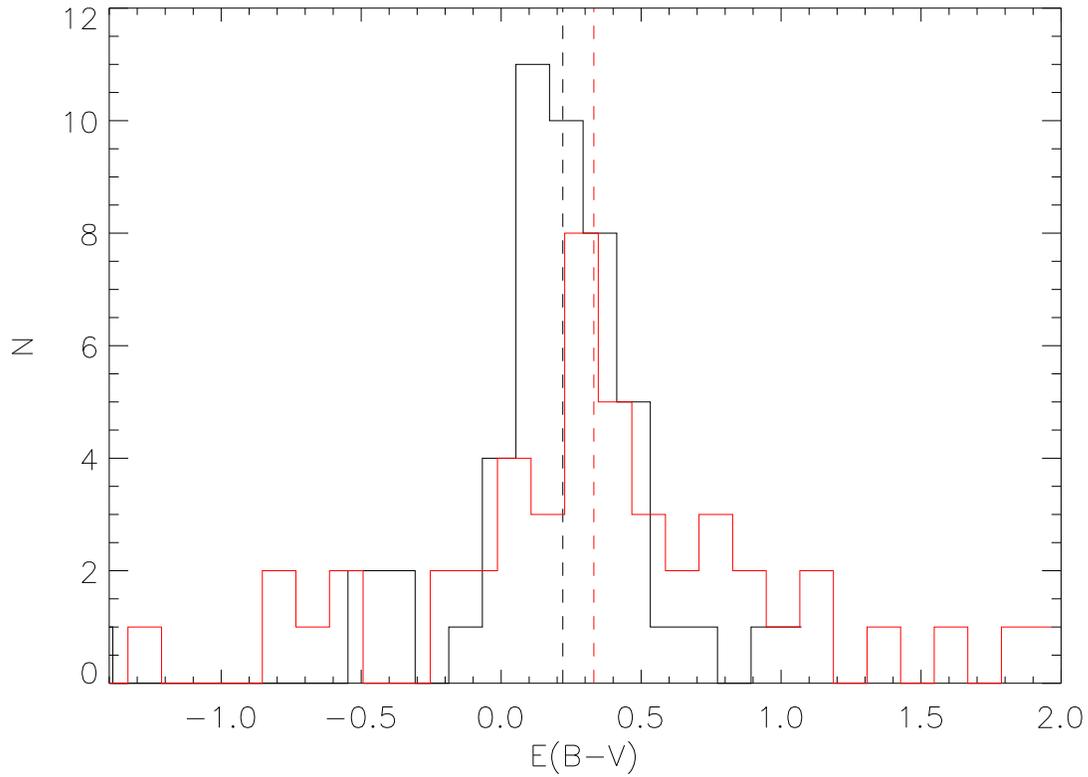}
\figcaption{Histograms showing distribution of $E(B-V)$ values of
each star forming region. {\it Red line}: distribution based on the 
measured Balmer Decrements. {\it Black line}: distribution based on the 
difference between modeled colors and observed colors. Single 
Gaussian fits yield mean values of $E(B-V)$ equal to 0.39~mag and 
0.28~mag respectively, indicated by vertical dashed lines. The 2$\sigma$
width of the Gaussian function fitted to the distribution represented 
by the black line is 0.36~mag.}
\label{fig3}
\end{minipage}
\end{figure}

\clearpage

\begin{figure}
\figurenum{4a}
\begin{minipage}{6.25in}
\plotone{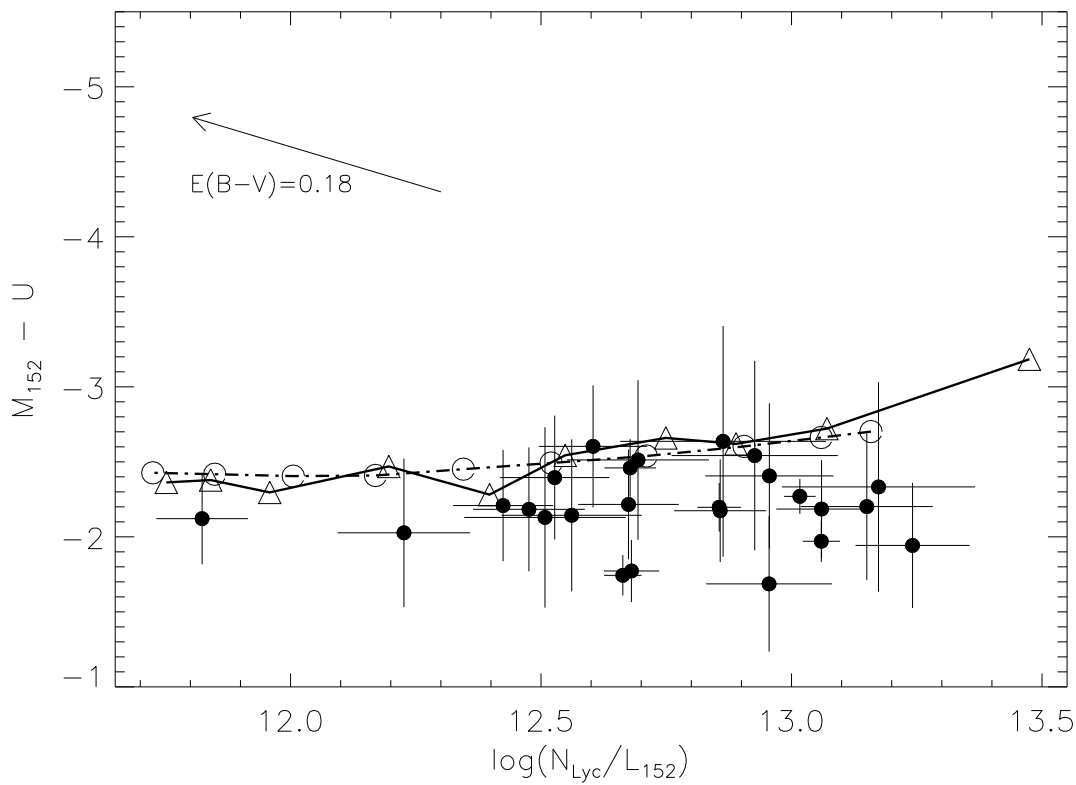}
\caption{Color vs. log$_{10}$(N$_{Lyc}$/L$_{152}$) diagrams. 
Colors M$_{152}$-$U$, M$_{152}$-$B$ and M$_{152}$-$V$ are plotted in  panels a,
b and c respectively. Error bars indicate 1$\sigma$ errors. Numbers in
panel c identify associated apertures. Filled circles: observational results
corrected for dust attenuation using the formalism of 
\protect\citet{cal94}.  
Open triangles: values derived from instantaneous burst models
computed at integer burst times beginning at 2 Myr on right and
increasing to 11 Myr on left. Open circles: values derived from
continuous star formation models computed at integer turn-off times
beginning at 0 Myr on right and increasing to 9 Myr on left. The arrow
indicates the effect of increasing the correction for extinction on
the observed colors arising from an increase in $E(B-V)$ of 0.18 mag.}
\label{fig4a}
\end{minipage}
\end{figure}

\clearpage

\begin{figure}
\figurenum{4b}
\begin{minipage}{6.25in}
\plotone{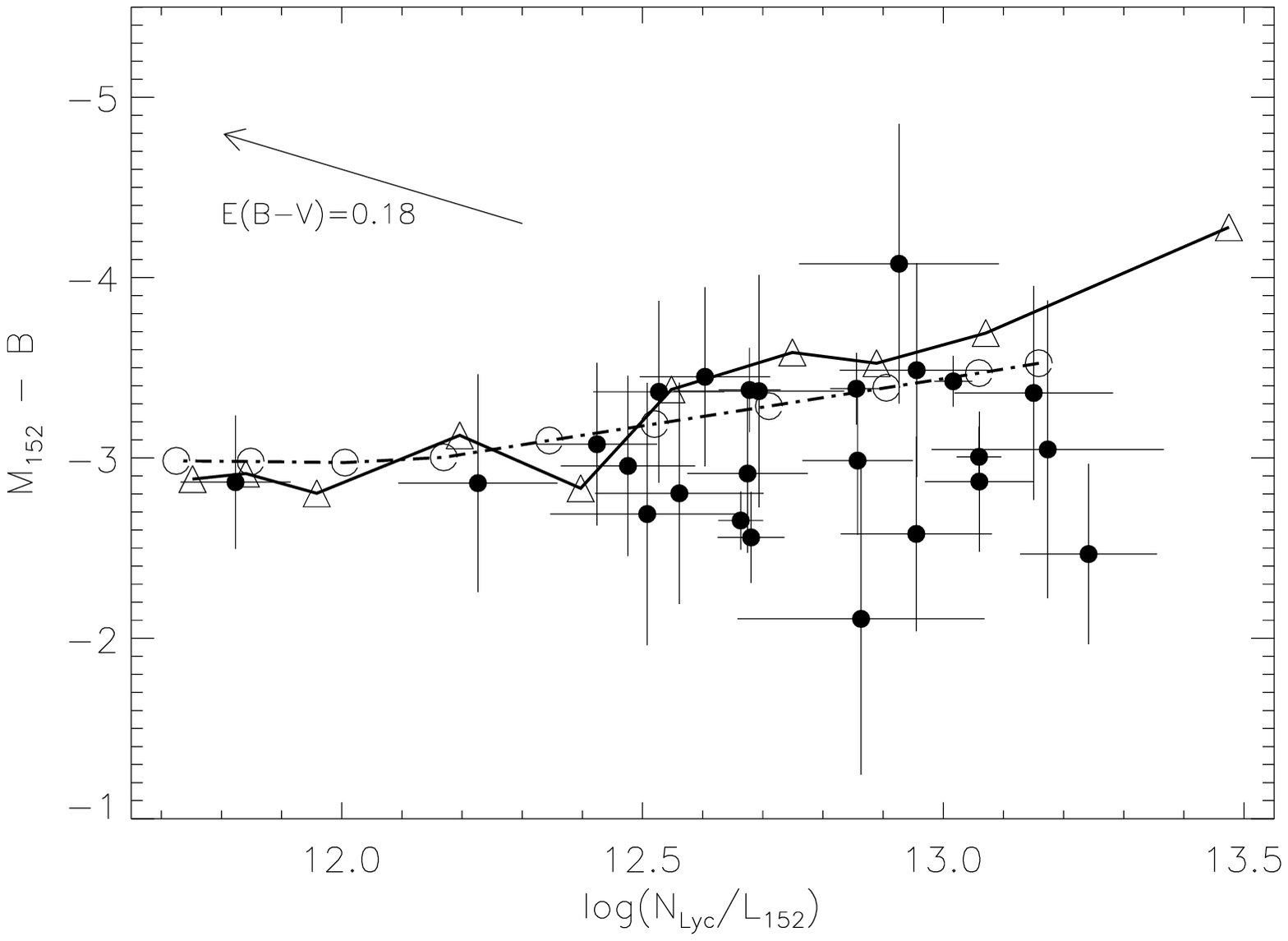}
\caption{M$_{152}$-$B$ vs. log$_{10}$(N$_{Lyc}$/L$_{152}$) diagram.}
\label{fig4b}
\end{minipage}
\end{figure}

\clearpage

\begin{figure}
\figurenum{4c}
\begin{minipage}{6.25in}
\plotone{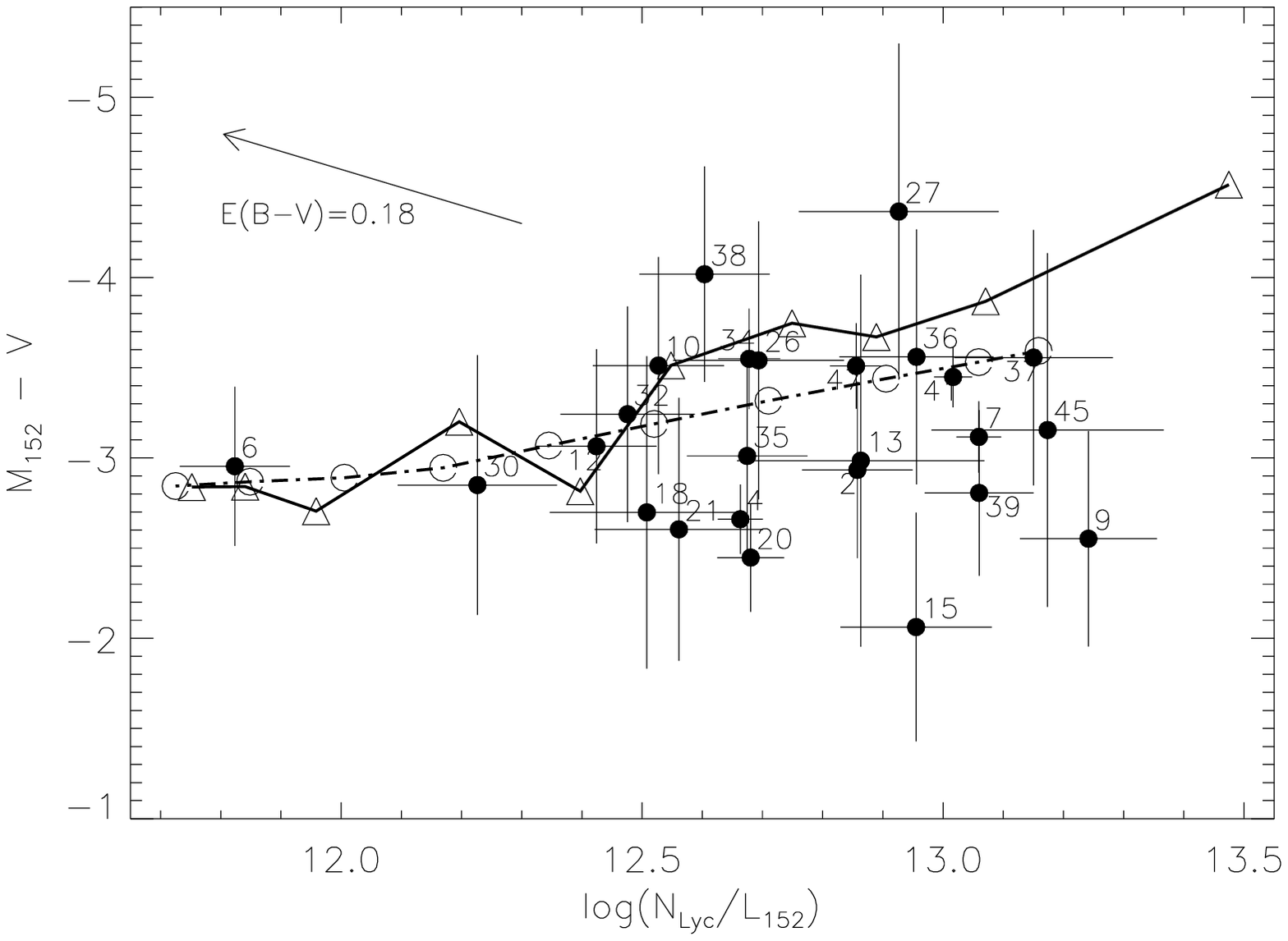}
\caption{M$_{152}$-$V$ vs. log$_{10}$(N$_{Lyc}$/L$_{152}$) diagram.}
\label{fig4c}
\end{minipage}
\end{figure}

\clearpage

%%%%%%%%%%%%%%%%%%%%%%%%%%%%%%%%%%%%%%%%%%%%%%%%%%%%%%%%%%%%%%%%%%%%%%%%%%%%%%

\begin{deluxetable}{lrrcr}
\tablecolumns{5}
%\tabletypesize{\scriptsize}
\tablewidth{0pc}
\tablenum{1}
\tablecaption{Observations\label{tab:1}}
\tablehead{
\colhead{Instrument} & 
\colhead{Date} & 
\colhead{Exposure} & 
\colhead{Bandpass} & 
\colhead{Resolution}\\
\colhead{} & 
\colhead{} & 
\colhead{(seconds)} & 
\colhead{$\lambda_{0} / \Delta\lambda$} & 
\colhead{(FWHM)}}

\startdata

UIT         &  03/10/1995   &   1140 &  FUV, 152.1/35.4 nm       &    3.4\arcsec \\     
& & & & \\
Kitt Peak 0.9m  &  05/16/1996   &    600 &         U      &    3.0\arcsec \\
            &  05/16/1996   &    300 &         B      &    1.6\arcsec \\
            &  05/16/1996   &    300 &         V      &    1.1\arcsec \\
            &  05/16/1996   &    180 &         R      &    1.6\arcsec \\
& & & & \\
Palomar\tablenotemark{a}  1.5m & 05/1990      &   2000 &  H$\alpha$, 658.5/1.5 nm  & 1.2\arcsec \\
& & & & \\
Mt. Laguna  &  05/23/1996     &    600 & H$\beta$, 487.8/5.0 nm      &    4.5\arcsec \\

\tablenotetext{a}{The H$\alpha$ data was provided by Dr. R. J. Rand. See Rand et al. (1992).}

\enddata

\end{deluxetable}

%%%%%%%%%%%%%%%%%%%%%%%%%%%%%%%%%%%%%%%%%%%%%%%%%%%%%%%%%%%%%%%%%%%%%%%%%%%%%%

\begin{deluxetable}{lccrrr}
\tablecolumns{6}
\tablewidth{0pc}
\small
\tablenum{4}
\tablecaption{FUV and IR Luminosities of Galaxies\label{tab:4}}
\tablehead{
\colhead{Name} & 
\colhead{AB\tablenotemark{a}} & 
\colhead{Distance\tablenotemark{b}} & 
\colhead{L$_{FUV}$\tablenotemark{c}} & 
\colhead{L$_{IR}$\tablenotemark{d}} &
\colhead{L$_{IR}$/L$_{FUV}$}  \\
\colhead{} & 
\colhead{(Mag)} & 
\colhead{(Mpc)} & 
\colhead{L$_{\sun}$} &
\colhead{L$_{\sun}$} & 
\colhead{}}

\startdata

   NGC~4631        & 0.02 &  7.5   &  6.99$\times$10$^{8}$ &  9.34$\times$10$^{9}$ &  13.40 \\
   NGC~5236 (M83)  & 0.14 &  4.8   &  8.68$\times$10$^{8}$ &  1.20$\times$10$^{10}$ &  13.80 \\
   NGC~5457 (M101) & 0.12 &  7.7   &  2.77$\times$10$^{9}$ &  1.12$\times$10$^{10}$ &   4.04 \\
   NGC~628  (M74)  & 0.12 &  7.8   &  3.34$\times$10$^{8}$ &  2.85$\times$10$^{9}$ &   8.53 \\
   NGC~5194 (M51)  & 0.00 &  8.4   &  1.23$\times$10$^{9}$ &  1.08$\times$10$^{10}$ &   8.78 \\

\tablenotetext{a}{Foreground Galactic extinction: NGC~4631, NGC~~628, NGC~5194, Burstein and Heiles 1984;
                  NGC~5236, AND NGC~5457, Hill {\it et al.}(1994).}
\tablenotetext{b}{NGC~5236, Pierce (1993); NGC~5457 and NGC~5194, Feldmeir and Ciardullo (1997); 
                  NGC~628, Sharina (1996).}
\tablenotetext{c}{Except for NGC~4631, luminosities  are computed from Waller 
                      (1997b), and are  adjusted for the distances given in Table~4.}
\tablenotetext{d}{IRAS luminosities are from (Rice {\it et al.} 1988) and are  
                   adjusted for distances given in Table~4.}

\enddata

\end{deluxetable}

%%%%%%%%%%%%%%%%%%%%%%%%%%%%%%%%%%%%%%%%%%%%%%%%%%%%%%%%%%%%%%%%%%%%%%%%%%%%%%

\end{document}